\documentstyle[12pt]{article}
\begin{document}
\hfill{SMU-HEP 94-25}
\vglue 0.1cm
\hfill{ITP-SB-94-59}
\vglue 0.1cm
\hfill{INLO-PUB-16/94}
\vskip 0.1cm
\vskip 0.2cm
\centerline{\large\bf { Rates for Inclusive Deep-Inelastic
Electroproduction}}
\vskip 0.2cm
\centerline{\large\bf {of Charm Quarks at HERA}}
\vskip 0.3cm
\centerline {\sc S. Riemersma}
\vskip 0.3cm
\centerline{\it Department of Physics,}
\centerline{\it Fondren Science Building,}
\centerline{\it Southern Methodist University,}
\centerline{\it Dallas, Texas 75275}
\vskip 0.3cm
\centerline {\sc J. Smith }
\vskip 0.3cm
\centerline{\it Institute for Theoretical Physics,}
\centerline{\it State University of New York at Stony Brook,}
\centerline{\it Stony Brook, New York 11794-3840}
\vskip 0.3cm
\centerline{and}
\vskip 0.3cm
\centerline{\sc W. L. van Neerven}
\vskip 0.3cm
\centerline{\it Instituut Lorentz,}
\centerline{\it University of Leiden,}
\centerline{\it P.O.B. 9506, 2300 RA, Leiden,}
\centerline{\it The Netherlands.}
\vskip 0.3cm
\centerline{November 1994}
\vskip 0.3cm
\centerline{\bf Abstract}
\vskip 0.4cm

The coefficient functions for heavy-flavour production in
deeply inelastic electron-hadron scattering have been calculated previously.
Analytic expressions are impossible to publish due to their length.
Therefore we have tabulated
them as two-dimensional arrays as is often
done for the scale-dependent parton densities. Using this
computer program we
present event rates for charm production at HERA in bins of
$x$ and $Q^2$.  These rates are insensitive to variations in the
factorization and renormalization scale $\mu$.
\vfill
\newpage
In the past few years, calculations of
$O(\alpha_s)$ QCD corrections to heavy-flavour production have made great
progress (for a recent review see \cite{st}).
Calculations have been completed for
hadron-hadron collisions
\cite{nde},\cite{bkns}, photoproduction \cite{en},\cite{sn},
electroproduction \cite{lrsn1}, \cite{aiv}, and
photon-photon collisions (real as well as virtual photons)
\cite{dkzz}, \cite{lrsn2}. The calculation of higher order corrections to
these processes are important for the the top quark search
\cite{cdf} and the determination of the gluon distribution
function \cite{aw}, \cite{grs} which can be measured in open charm production.

The expressions are only available in
large computer programs
for the radiative corrections as their complexity prohibits publishing
them in an analytic form. This complexity is
due to the non-zero mass $m$ of the heavy quark. If
$m$ were zero the final formulae could be constructed and
published as is shown e.g., for the $O(\alpha_s^2)$ corrections to the
coefficient functions in deeply inelastic lepton-hadron scattering
\cite{zn} and the Drell-Yan process \cite{hnm}.
In some special cases the use of lengthy computer programs
can be avoided by either making approximations \cite{mssn} or
by making algebraic fits to the exact coefficient functions
of heavy-flavour production in hadron-hadron collisions \cite{nde} and
in photon-hadron scattering \cite{sn}. Such fits are quite accurate and
enabled the authors in \cite{admn} to present complete tables of
cross sections for charm, bottom and top production in many hadron-hadron
reactions.

These algebraic fits mentioned above could be made because the coefficient
functions calculated for
heavy-flavour production in hadron-hadron and photon-hadron
collisions only depend on one scale independent variable.
This is in contrast with the coefficient functions of deeply inelastic
heavy-flavour production which depend on two scale independent
variables as the photon is virtual. Moreover it
turns out that in the latter process the coefficient functions show a much
more complicated behaviour than in the former ones so that an
algebraic fit as presented in \cite{nde} is very difficult to
achieve. Therefore we have constructed a set of tables reproducing
the coefficient functions for each partonic subprocess in deeply inelastic
electroproduction of heavy flavours. These tables are presented in the form of
a two dimensional array in a computer program, analogous to the
way various groups \cite{cteq}, \cite{msr} present the scale
dependent parton densities in hadrons, which also
depend on two variables. The computer program is available.
\footnote{Requests should be sent to smith@elsebeth.physics.sunysb.edu.}
Our two-dimensional tables are used to predict the $O(\alpha_s)$ corrected
rates for inclusive charm production at HERA by adding integrations
over bins in $x$ and $Q^2$.
It will be interesting to compare our results with
the forthcoming data from the ZEUS and H1 collaborations.
Notice that in deeply inelastic electroproduction
we only have to contend with the parton densities in the proton.
In a later paper we will address the complications associated
with predicting rates for heavy-flavour
photoproduction (small $Q^2$) which also involve
the parton densities in the photon.

We begin by listing several important formulae.
Omitting charged-current interactions, heavy-flavour
production in deeply inelastic electron-proton scattering proceeds via
the reaction
\begin{equation}
e^-(l_1) + P(p) \rightarrow e^-(l_2) + Q(p_1)(\bar Q(p_2)) + X\,.
\end{equation}
Here $X$ stands for any hadronic final state allowed by quantum number
conservation. We sum over these states so that the process is
inclusive with respect to the outgoing hadrons. When the virtuality
$-q^2$  $(q=l_1 - l_2)$ of the exchanged photon
is not too large $(-q^2 \ll M_Z^2)$ the reaction in (1) is
dominated by the one-photon exchange mechanism and we can neglect any weak
interaction effects.
If we also integrate over the heavy (anti)quark $Q$ $(\bar Q)$ in the
final state the deeply inelastic electroproduction cross section
can be written as
\begin{equation}
\frac{d^2\sigma}{dxdy} = \frac{2\pi\alpha^2}{Q^4} S
[\{1+(1-y)^2\}F_2(x,Q^2,m^2)-y^2F_L(x,Q^2,m^2)] \,,
\end{equation}
where $S $ denotes the square of the c.m. energy of the
electron-proton system. The variables $x$ and $y$ are defined
as
\begin{equation}
x = \frac{Q^2}{2p\cdot q} \quad (0 < x \le 1)\quad ,\quad
y = \frac{p\cdot q}{p\cdot l_1} \quad (0<y < 1) \,,
\end{equation}
with
\begin{equation}
 -q^2 = Q^2 = xyS \,.
\end{equation}
The deeply
inelastic heavy-flavour structure functions appearing in the cross
section (2) are given by $F_2(x,Q^2,m^2)$ and $F_L(x,Q^2,m^2)$
(longitudinal). The structure functions are given by the formula
(see (6.5) of \cite{lrsn1})
\begin{eqnarray}
F_{k}(x,Q^2,m^2) &=&
\frac{Q^2 \alpha_s}{4\pi^2 m^2}
\int_x^{z_{\rm max}} \frac{dz}{z}  \Big[ \,e_H^2 f_g(\frac{x}{z},\mu^2)
 c^{(0)}_{k,g} \,\Big] \nonumber \\&&
+\frac{Q^2 \alpha_s^2}{\pi m^2}
\int_x^{z_{\rm max}} \frac{dz}{z}  \Big[ \,e_H^2 f_g(\frac{x}{z},\mu^2)
 (c^{(1)}_{k,g} + \bar c^{(1)}_{k,g} \ln \frac{\mu^2}{m^2}) \nonumber \\ &&
+\sum_{i=q,\bar q} \Big[ e_H^2\,f_i(\frac{x}{z},\mu^2)
 (c^{(1)}_{k,i} + \bar c^{(1)}_{k,i} \ln \frac{\mu^2}{m^2}) \nonumber \\ &&
+ e^2_{L,i}\, f_i(\frac{x}{z},\mu^2) (d^{(1)}_{k,i} + \bar d^{(1)}_{k,i}
\ln\frac{\mu^2}{m^2}) \, \Big]  \,\Big] \,,
\end{eqnarray}
where $k = 2,L$ and the upper boundary on the integration is given by
$z_{\rm max} = Q^2/(Q^2+4m^2)$. Further $f_i(x,\mu^2)\,, (i=g,q,\bar q)$
denote the parton densities in the proton and $\mu$ stands for the
mass factorization scale,
which has been put equal to the renormalization scale. The coefficient
functions, represented by $c^{(l)}_{k,i}(\eta, \xi)\,,\bar c^{(l)}_{k,i}
(\eta, \xi)\,,
(i=g\,,q\,,\bar q\,;l=0,1)$
and by $d^{(l)}_{k,i}(\eta, \xi)\,,\bar d^{(l)}_{k,i}(\eta, \xi)$,
$(i=q\,,\bar q\,;l=0,1)$
are calculated in \cite{lrsn1} and they are represented in the
$\overline{\rm MS}$ scheme.
Furthermore they depend on the scaling variables $\eta$ and $\xi$ defined by
\begin{equation}
\eta = \frac{s}{4m^2} - 1\quad \,, \qquad \xi = \frac{Q^2}{m^2}\,.
\end{equation}
where $s$ is the square of the c.m. energy of the
virtual photon-parton subprocess
which implies that in (5) $z=Q^2/(Q^2+s)$. In this equation we made
a distinction between the coefficient functions with respect to their origin.
The coefficient functions indicated by
$c^{(l)}_{k,i}(\eta, \xi),\bar c^{(l)}_{k,i}(\eta, \xi)$
originate from the partonic
subprocesses where the virtual photon is coupled to the heavy quark whereas
the quantities $d^{(l)}_{k,i}(\eta, \xi)\,,\bar d^{(l)}_{k,i}(\eta, \xi)$
come from the subprocess where the virtual
photon interacts with the light quark.
Hence the former are multiplied by the charge squared
of the heavy quark $e_H^2$, whereas the latter are
multiplied by the charge squared of the light quark $e_L^2$ respectively
(both in units of $e$).
Although terms proportional to $e_H e_L$ appear in the
inclusive photon-parton differential distributions they integrate
to zero in the total partonic cross section, so we have not included them
in (6).
Furthermore we have isolated the factorization scale dependent term
$\ln(\mu^2/m^2)$. The functions multiplied by this term, which are
indicated by a bar, are called mass factorization parts.
Notice that in the subsequent equations
we discuss the transverse coefficient functions indicated by the
subscript $T$ instead of the ones indicated by the subscript $2$.
The relation between them is given by
$c^{(l)}_{2,i}(\eta,\xi) = c^{(l)}_{T,i}(\eta,\xi) + c^{(l)}_{L,i}(\eta,\xi)$
and
$d^{(l)}_{2,i}(\eta,\xi) = d^{(l)}_{T,i}(\eta,\xi) + d^{(l)}_{L,i}(\eta,\xi)$.
where the same definition holds for the coefficient functions indicated by
a bar.  In the limit $\xi\rightarrow 0$ (see (6)) where the virtual photon
becomes
on-shell the above coefficient functions tend to their analogues obtained for
photoproduction which can be found in \cite{en},\cite{sn}. For that purpose
we had to modify the original expressions for the functions
$d^{(1)}_{k,q}(\eta,\xi)$.Originally $\bar d^{(1)}_{k,q}(\eta,\xi)$ did not
exist because no mass factorization was needed and in the limit
$\xi\rightarrow 0$ $d^{(1)}_{T,q}(\eta,\xi)$ diverged logarithmically. This
is due to an additional collinear divergence which appears when the virtual
photon coupled to the light quark goes on-mass-shell.
Furthermore in the same limit $d^{(1)}_{L,q}(\eta,\xi)$ did not vanish.
Therefore, to use these functions in the region
$\xi \approx 0$ one has to perform an additional mass factorization to remove
the collinear divergence which is due to the on-shell photon. This is achieved
by subtracting a term which is multiplied by a scale invariant function
called $R(\xi) = {\rm exp}(-20 \xi)$ ( see (5.10) in \cite{lrsn1} ) where the
subtraction is imposed if $Q^2 < Q^2_{min} = 1.5$ GeV$^2$.
This leads to the appearance of the function $\bar d^{(1)}_{T,q}$
 ( $\bar d^{(1)}_{L,q} = 0$ ) which would
not have been present when the photon is treated to be highly virtual.
This implies that the function $\bar d^{(1)}_{T,q}$ will be proportional to
$R(\xi)$ and it vanishes when $\xi\rightarrow \infty$.
Notice that our choice of $R$
in \cite{lrsn1} was not scale independent so that the plot 11.b in that
paper still contains a scale dependence. This however has no consequence for
any of the numerical results. The above procedure implies that
in the on-mass-shell limit the function $d^{(1)}_{L,q}(\eta,0) = 0$
and $d^{(1)}_{T,q}(\eta,0)$,$\bar d^{(1)}_{T,q}(\eta,0)$ become equal
to the on-mass-shell
photon coefficient functions in \cite{en}, \cite{sn} (see (2.11) and (2.15)
in \cite{en}).

The coefficient functions for the Born reaction (virtual photon-gluon fusion,
see \cite{ew}) are given by
\begin{eqnarray}
c^{(0)}_{L,g}(\eta,\xi) &=& \frac{\pi}{2} T_f \frac{\xi}{(1+\eta+\xi/4)^3}
\Big[2 ({\eta(1+\eta)})^{1/2} -\ln\frac
{(1+\eta)^{1/2}+\eta^{1/2}}
{(1+\eta)^{1/2}-\eta^{1/2}} \Big]\,,
\nonumber \\
\end{eqnarray}
\begin{eqnarray}
c^{(0)}_{T,g}(\eta,\xi) &=& \frac{\pi}{2} T_f \frac{1}{(1+\eta+\xi/4)^3}
\Big[-2 \Big\{(1+\eta-\xi/4)^2 +1 + \eta\Big\}
\Big(\frac{\eta}{1+\eta}\Big)^{1/2} \nonumber \\ &&
+\Big\{ 2(1+\eta)^2 +\frac{\xi^2}{8} + 1 + 2 \eta \Big\}\ln\frac
{(1+\eta)^{1/2}+\eta^{1/2}}
{(1+\eta)^{1/2}-\eta^{1/2}} \Big]\,,
\end{eqnarray}
where $T_f=1/2$ in $SU(N)$.

No such simple analytic expressions
can be given for the next to leading order coefficient functions.
Therefore we present them in tables constructed as follows.
First we ran the programs in \cite{lrsn1}
and computed the coefficient functions for a grid
of values of $\eta$ and $\xi$ as defined in (6).This we have only done when
they are represented in the $\overline{\rm MS}$ scheme.
We then divided these functions by the
appropriate colour factor and then subtracted the asymptotic
and threshold dependences,
for which analytic expressions are available in the literature and will be
presented below. Finally we wrote subroutines to set up two dimensional
arrays. The interpolation is done in a bilinear fashion \cite{pt}.

Starting with the virtual-gluon subprocess we define a new function with the
threshold and asymptotic behavior removed, namely
\begin{eqnarray}
&& h^{(1)}_{A,k,g}(\eta, \xi) = (C_A T_f)^{-1}\, c^{(1)}_{A,k,g}(\eta,\xi) -
\beta G_k(\eta,\xi) - \rho E_{k,A}(\eta,\xi) \, ,
\end{eqnarray}
and
\begin{eqnarray}
&& h^{(1)}_{F,k,g}(\eta, \xi) = (C_AT_f)^{-1}\, c^{(1)}_{F,k,g}(\eta, \xi) -
\rho E_{k,F}(\eta,\xi) \, .
\end{eqnarray}
Here we have split the coefficient functions $c^{(1)}_{k,g}$ according to
their colour parts indicated by the subscripts $A$ and $F$.
The colour factors are given by $C_A T_f$ and $C_F T_f$
respectively, where for $SU(N)$ , $C_A = N$ and $C_F = (N^2 - 1)/2 N$.
Further we have defined
\begin{equation}
\beta = \Big(\frac{\eta}{1+\eta}\Big)^{1/2}
\quad\,, \qquad \rho = \frac{1}{1+\eta}\,.
\end{equation}

The mass factorization parts $\bar c_{k,g}(\eta,\xi)$ can be parameterized
in a similar way by
\begin{eqnarray}
&& \bar h^{(1)}_{k,g}(\eta, \xi) = (C_A T_f)^{-1}\,
\bar c^{(1)}_{k,g}(\eta, \xi)
 - \beta \bar G_k(\eta,\xi) - \rho \bar E_{k,A}(\eta,\xi).
\end{eqnarray}

The functions $E_{k,C},\bar E_{k,C}$ with $k=T\,,L$
and $C=A\,,F$ describe the threshold behaviour as
$\eta \rightarrow 0$ (or as $s\rightarrow 4m^2$) and are derived from
(5.7)-(5.9) of \cite{lrsn1}
\footnote{Notice that an extra factor of two
should be multiplied to (5.9) of \cite{lrsn1}} .
The asymptotic behavior which holds in the region
$\eta\rightarrow \infty $ (or as $(s\rightarrow \infty)$ is
given by the functions
$G_k,\bar G_k$ with $k=T\,,L$. The latter are obtained from
\cite{cch} (see their Appendix A).
The functions describing the threshold region have the following form
\begin{equation}
E_{L,F}(\eta,\xi) = \frac{1}{6\pi}\frac{\xi}{(1+\xi/4)^3}\,\beta^2
\,\Big[\frac{\pi^2}{2}\Big]\,,
\end{equation}
\begin{equation}
E_{T,F}(\eta,\xi) = \frac{1}{4\pi}\frac{1}{1+\xi/4}
\,\Big[\frac{\pi^2}{2}\Big]\,,
\end{equation}
\begin{equation}
E_{L,A}(\eta,\xi) = \frac{1}{6\pi}\frac{\xi}{(1+\xi/4)^3}\,\beta^2
\Big[ \, \beta\ln^2(8\beta^2) - 5\beta\ln(8\beta^2) - \frac{\pi^2}{4} \, \Big]
\,,
\end{equation}
\begin{equation}
E_{T,A}(\eta,\xi) = \frac{1}{4\pi}\frac{1}{(1+\xi/4)}\,
\Big[\beta\ln^2(8\beta^2) - 5\beta\ln(8\beta^2) - \frac{\pi^2}{4}\Big]\,,
\end{equation}
\begin{equation}
\bar E_{L,A}(\eta,\xi) = \frac{1}{6\pi}\frac{\xi}{(1+\xi/4)^3}\,\beta^3
\,\Big[-\ln(4\beta^2)\Big]\,,
\end{equation}
\begin{equation}
\bar E_{T,A}(\eta,\xi) = \frac{1}{4\pi}\frac{1}{1+\xi/4}\,\beta
\,\Big[-\ln(4\beta^2)\Big]\,.
\end{equation}
The functions describing the asymptotic region are given by
\begin{eqnarray}
G_L(\eta,\xi)&=&\frac{1}{6\pi}\Big[ \frac{4}{\xi}
- \frac{4}{3} \frac{1}{1+\xi/4} + (1-\frac{2}{\xi}
-\frac{1}{6}\frac{1}{1+\xi/4} ) J(\xi) \nonumber \\ &&
-\Big( \frac{3}{\xi} +\frac{1}{4} \frac{1}{1+\xi/4}\Big) I(\xi)\Big] \,,
\end{eqnarray}
\begin{eqnarray}
G_T(\eta,\xi)&=&\frac{1}{6\pi}\Big[ -\frac{2}{3}\frac{1}{\xi}
+ \frac{4}{3} \frac{1}{1+\xi/4} + (\frac{7}{6}+\frac{1}{3}\frac{1}{\xi}
+\frac{1}{6}\frac{1}{1+\xi/4} ) J(\xi) \nonumber \\ &&
+\Big(1+\frac{2}{\xi} +\frac{1}{4}\frac{1}{1+\xi/4}\Big) I(\xi)\Big] \,,
\end{eqnarray}
\begin{eqnarray}
\bar G_L(\eta,\xi)&=&\frac{1}{6\pi}\Big[-\frac{6}{\xi}
+ \frac{1}{2} \frac{1}{1+\xi/4} + (\frac{3}{\xi}
+\frac{1}{4}\frac{1}{1+\xi/4} ) J(\xi) \Big]\,,
\end{eqnarray}
\begin{eqnarray}
\bar G_T(\eta,\xi)&=&\frac{1}{6\pi}\Big[\frac{4}{\xi}
- \frac{1}{2} \frac{1}{1+\xi/4} - ( 1 + \frac{2}{\xi}
+\frac{1}{4}\frac{1}{1+\xi/4} ) J(\xi) \Big]\,,
\end{eqnarray}
where the functions $J(\xi)$ and $I(\xi)$ are defined by
(see Appendix A in \cite{cch})
\begin{eqnarray}
J(\xi) = \frac{4}{(\xi(4+\xi))^{1/2}}
\ln\Big(\frac{(4+\xi)^{1/2} + \xi^{1/2}}
{(4+\xi)^{1/2} - \xi^{1/2}} \Big)\,,
\end{eqnarray}
\begin{eqnarray}
I(\xi) &=& \frac{4}{(\xi(4+\xi))^{1/2}} \Big[
-\frac{\pi^2}{6}- \frac{1}{2}
\ln^2\Big(\frac{(4+\xi)^{1/2} + \xi^{1/2}}
{(4+\xi)^{1/2} - \xi^{1/2}} \Big) \nonumber \\ &&
+ \ln^2 \Big(\frac{(4+\xi)^{1/2} - \xi^{1/2}}
{2(4+\xi)^{1/2}}\Big)
+ 2 {\rm Li}_2 \Big(\frac{(4+\xi)^{1/2} - \xi^{1/2}}
{2(4+\xi)^{1/2}} \Big)\Big]\,,
\end{eqnarray}
where ${\rm Li}_2(x)$ is the dilogarithmic function defined as
\begin{eqnarray}
{\rm Li}_2(x) = - \int_0^x \frac{dt}{t} \, \ln(1-t)\,.
\end{eqnarray}
We proceed in an analogous way for the coefficient functions corresponding
to the virtual photon-light quark subprocesses.
When the photon is coupled to the heavy flavour they
are parameterized as
\begin{eqnarray}
&& h^{(1)}_{H,k,q}(\eta, \xi) = (C_F T_f)^{-1}\,c^{(1)}_{k,q}(\eta, \xi) -
\beta^3 G_k(\eta,\xi),
\end{eqnarray}
and
\begin{eqnarray}
&& \bar h^{(1)}_{H,k,q}(\eta, \xi) =
(C_F T_f)^{-1}\,\bar c^{(1)}_{k,q}(\eta, \xi)
 - \beta^3 \bar G_k(\eta,\xi)\,.
\end{eqnarray}
When the photon is coupled to the light quark we get
\begin{eqnarray}
&& h^{(1)}_{L,k,q}(\eta, \xi) = (C_F T_f)^{-1} d^{(1)}_{k,q}(\eta, \xi)\, ,
\end{eqnarray}
and
\begin{eqnarray}
&& \bar h^{(1)}_{L,k,q}(\eta, \xi) =
 (C_F T_f)^{-1}\,\bar d^{(1)}_{k,q}(\eta, \xi) \,.
\end{eqnarray}
The subscripts $H$ and $L$ in the above expressions indicate that when
they are inserted in (5) they have to be multiplied by the charge factors
$e_H^2$ and $e_L^2$ respectively.

Notice that for the reactions discussed above only the $A$-part
of the coefficient
functions show an enhancement in both the threshold region and the
asymptotic region. For the $F$-part we only observe large corrections in the
threshold region except for the process given by the expressions in (28)
and (29).
To illustrate the quality of the fits, we present the plots of the coefficient
functions which constitute the bulk of the $O(\alpha_s)$
radiative correction. They are given by
$h^{(1)}_{A,T,g}$ (9) and $h^{(1)}_{F,T,g}$ (10), which are shown in
fig.1 and fig.2 respectively . They are plotted versus $\eta$ for
several different values of $\xi$. (One can compare them with the
plots in \cite{lrsn1} for the corresponding
functions without the subtraction of the
threshold and asymptotic behavior.) In both instances, we observe
that the $h$ functions tend to zero as $ \eta \rightarrow 0$
and as $\eta \rightarrow \infty$.
Also, as $\xi \rightarrow \infty $, the $h$ functions become zero
across the entire $\eta$-region.
We also see the complicated behaviour of the functions
in the intermediate region which illustrates why an algebraic
parameterization is very difficult to accomplish.

Using the fits, we now present single-particle inclusive event
rates for inclusive $c$ production given an
integrated luminosity of 300 nb$^{-1}$ at $ \sqrt{S} = 298 $ GeV.
Notice that we consider charm production only and do not sum over the charm
and anti-charm cross sections.
We take $m_c = 1.6$ GeV${/c^2}$ and vary the mass factorization
scale $\mu  = \sqrt{Q^2 + m^2}$
up and down by a factor of two.  Furthermore we have chosen the CTEQ2M parton
densities \cite{cteq} and the two-loop running coupling constant with
$\Lambda_{\rm QCD}^{(4)} = 213$ MeV. Our results are listed in tables 1 and 2
where we computed the cross section in (2) in bins of $x$ and $Q^2$
(these are the bins used by the ZEUS collaboration in their
1993 data for $F_2$).
Table 1 contains the event numbers for the
lower $Q^2$ bins and Table 2 the corresponding numbers for the
higher $Q^2$ bins.

We find the events concentrated at low $Q^2$ and $x$, with an approximate
five percent uncertainty at small $x$ coming from the variation in the
scale $\mu$. In the
intermediate $Q^2$ and $x$ region, we find the number of events stable
and significant.  As $Q^2$ increases, the number of events drops rapidly.
However, here we expect weak interaction effects to reduce
the applicability of the
one-photon exchange approximation.  To conclude, we see that the number of
charm events is large and relatively insusceptible to variations in
the scale.
{}From these results and the good luminosity at HERA,
the extraction of the gluon density with significantly reduced uncertainty
should be possible.
\vskip 0.5 cm
\centerline{\bf Acknowledgements}
We thank J. Whitmore for useful discussions.  S.Riemersma thanks the ITP
at Stony Brook for their hospitality and the Lightner-Sams
Foundation, Inc. for assistance in providing computing facilities.
The research of J. Smith is supported in part by the contract
NSF 9309888.

\begin{tabular}{||c|c|c|c|c|c|c||} \hline \hline
$Q^2$     &  $Q^2$  &  $x$  &  $x$   & \multicolumn{3}{c||}{Events}  \\ \hline
(GeV$^2$) &  range  &       &  range & $\mu = M/2$ & $\mu = M$ & $\mu = M$ \\
\hline
 8.5      & 7 - 10  &  $4.2 \cdot 10^{-4}$ & $ 3.0 - 6.0 \cdot 10^{-4}$
& 228 & 205 & 204 \\ \cline{3-7}
          &         &  $8.5 \cdot 10^{-4}$ & $ 6.0 - 12.0 \cdot 10^{-4}$
& 196 & 176 & 173 \\ \hline
 12       & 10 - 14 &  $4.2 \cdot 10^{-4}$ & $ 3.0 - 6.0 \cdot 10^{-4}$
& 182 & 169 & 170 \\ \cline{3-7}
          &         &  $8.5 \cdot 10^{-4}$ & $ 6.0 - 12.0 \cdot 10^{-4}$
& 164 & 152 & 151 \\ \cline{3-7}
          &         &  $1.6 \cdot 10^{-3}$ & $ 1.2 - 2.0 \cdot 10^{-3}$
& 102 & 94  & 93  \\ \cline{3-7}
          &         &  $2.7 \cdot 10^{-3}$ & $ 2.0 - 3.6 \cdot 10^{-3}$
& 95 & 88  & 86  \\ \hline
 15       & 14 - 20 &  $4.2 \cdot 10^{-4}$ & $ 3.0 - 6.0 \cdot 10^{-4}$
& 153 & 143 & 144 \\ \cline{3-7}
          &         &  $8.5 \cdot 10^{-4}$ & $ 6.0 - 12.0 \cdot 10^{-4}$
& 148 & 137 & 137 \\ \cline{3-7}
          &         &  $1.6 \cdot 10^{-3}$ & $ 1.2 - 2.0 \cdot 10^{-3}$
& 94 & 88  & 86  \\ \cline{3-7}
          &         &  $2.7 \cdot 10^{-3}$ & $ 2.0 - 3.6 \cdot 10^{-3}$
& 90 & 83  & 81  \\ \cline{3-7}
          &         &  $4.7 \cdot 10^{-3}$ & $ 3.6 - 6.0 \cdot 10^{-3}$
& 62 & 57  & 55  \\ \hline
 25       & 20 - 28 &  $4.2 \cdot 10^{-4}$ & $ 3.0 - 6.0 \cdot 10^{-4}$
& 104 & 99 & 100 \\ \cline{3-7}
          &         &  $8.5 \cdot 10^{-4}$ & $ 6.0 - 12.0 \cdot 10^{-4}$
& 108 & 103 & 103 \\ \cline{3-7}
          &         &  $1.6 \cdot 10^{-3}$ & $ 1.2 - 2.0 \cdot 10^{-3}$
& 72 & 69  & 68  \\ \cline{3-7}
          &         &  $2.7 \cdot 10^{-3}$ & $ 2.0 - 3.6 \cdot 10^{-3}$
& 70 & 67  & 65  \\ \cline{3-7}
          &         &  $4.7 \cdot 10^{-3}$ & $ 3.6 - 6.0 \cdot 10^{-3}$
& 49 & 46  & 45  \\ \hline
35        & 28 - 40 &  $8.5 \cdot 10^{-4}$ & $ 6.0 - 12.0 \cdot 10^{-4}$
& 87 & 86 & 87 \\ \cline{3-7}
          &         &  $1.6 \cdot 10^{-3}$ & $ 1.2 - 2.0 \cdot 10^{-3}$
& 62 & 61  & 61  \\ \cline{3-7}
          &         &  $2.7 \cdot 10^{-3}$ & $ 2.0 - 3.6 \cdot 10^{-3}$
& 62 & 61 & 61  \\ \cline{3-7}
          &         &  $4.7 \cdot 10^{-3}$ & $ 3.6 - 6.0 \cdot 10^{-3}$
& 45 & 43  & 42  \\ \cline{3-7}
          &         &  $7.7 \cdot 10^{-3}$ & $ 6.0 - 10.0 \cdot 10^{-3}$
& 35 & 34  & 33  \\ \hline
50        & 40 - 56 &  $8.5 \cdot 10^{-4}$ & $ 6.0 - 12.0 \cdot 10^{-4}$
& 58 & 59 & 61 \\ \cline{3-7}
          &         &  $1.6 \cdot 10^{-3}$ & $ 1.2 - 2.0 \cdot 10^{-3}$
& 45 & 45  & 46  \\ \cline{3-7}
          &         &  $2.7 \cdot 10^{-3}$ & $ 2.0 - 3.6 \cdot 10^{-3}$
& 47 & 47  & 47  \\ \cline{3-7}
          &         &  $4.7 \cdot 10^{-3}$ & $ 3.6 - 6.0 \cdot 10^{-3}$
& 34 & 34  & 34  \\ \cline{3-7}
          &         &  $7.7 \cdot 10^{-3}$ & $ 6.0 - 10.0 \cdot 10^{-3}$
& 27 & 27  & 27  \\ \cline{3-7}
          &         &  $1.4 \cdot 10^{-2}$ & $ 1.0 - 2.0 \cdot 10^{-2}$
& 28 & 27  & 26  \\ \hline
65        & 56 - 80 &  $1.6 \cdot 10^{-3}$ & $ 1.2 - 2.0 \cdot 10^{-3}$
& 33 & 33  & 33  \\ \cline{3-7}
          &         &  $2.7 \cdot 10^{-3}$ & $ 2.0 - 3.6 \cdot 10^{-3}$
& 37 & 36  & 36  \\ \cline{3-7}
          &         &  $4.7 \cdot 10^{-3}$ & $ 3.6 - 6.0 \cdot 10^{-3}$
& 28 & 27  & 27  \\ \cline{3-7}
          &         &  $7.7 \cdot 10^{-3}$ & $ 6.0 - 10.0 \cdot 10^{-3}$
& 23 &22  & 22  \\ \cline{3-7}
          &         &  $1.4 \cdot 10^{-2}$ & $ 1.0 - 2.0 \cdot 10^{-2}$
& 23 & 22  & 21  \\ \hline
\end{tabular}
\newpage
\begin{tabular}{||c|c|c|c|c|c|c||} \hline \hline
$Q^2$     &  $Q^2$  &  $x$  &  $x$   & \multicolumn{3}{c||}{Events}  \\ \hline
(GeV$^2$) &  range  &       &  range & $\mu = M/2$ & $\mu = M$ & $\mu = M$ \\
\hline
 125      & 80 - 160& $1.6 \cdot 10^{-3}$ & $ 1.2 - 2.0 \cdot 10^{-3}$
& 28 & 28 & 28 \\ \cline{3-7}
          &         &  $2.7 \cdot 10^{-3}$ & $ 2.0 - 3.6 \cdot 10^{-3}$
& 41 & 41  & 41  \\ \cline{3-7}
          &         &  $4.7 \cdot 10^{-3}$ & $ 3.6 - 6.0 \cdot 10^{-3}$
& 33 & 33  & 33  \\ \cline{3-7}
          &         &  $7.7 \cdot 10^{-3}$ & $ 6.0 - 10.0 \cdot 10^{-3}$
& 29 &28  & 28  \\ \cline{3-7}
          &         &  $1.4 \cdot 10^{-2}$ & $ 1.0 - 2.0 \cdot 10^{-2}$
& 29 & 29  & 28  \\ \cline{3-7}
          &         &  $2.8 \cdot 10^{-2}$ & $ 2.0 - 4.0 \cdot 10^{-2}$
& 19 & 18 & 17 \\ \hline
 250      & 160 - 320& $4.7 \cdot 10^{-3}$ & $ 3.6 - 6.0 \cdot 10^{-3}$
& 15 & 15 & 15 \\ \cline{3-7}
          &         &  $7.7 \cdot 10^{-3}$ & $ 6.0 - 10.0 \cdot 10^{-3}$
& 15 & 14  & 14  \\ \cline{3-7}
          &         &  $1.4 \cdot 10^{-2}$ & $ 1.0 - 2.0 \cdot 10^{-2}$
& 16 & 16  & 15  \\ \cline{3-7}
          &         &  $2.8 \cdot 10^{-2}$ & $ 2.0 - 4.0 \cdot 10^{-2}$
& 11 & 10  & 10  \\ \cline{3-7}
          &         &  $5.7 \cdot 10^{-2}$ & $ 4.0 - 8.0 \cdot 10^{-2}$
& 6.2 & 5.8  & 5.3  \\ \hline
 500      & 320 - 640 &  $7.7 \cdot 10^{-3}$ & $ 6.0 - 10.0 \cdot 10^{-3}$
& 6.0 & 6.0  & 5.9  \\ \cline{3-7}
          &         &  $1.4 \cdot 10^{-2}$ & $ 1.0 - 2.0 \cdot 10^{-2}$
& 7.6 & 7.5  & 7.3  \\ \cline{3-7}
          &         &  $2.8 \cdot 10^{-2}$ & $ 2.0 - 4.0 \cdot 10^{-2}$
& 5.6 &5.4  & 5.2  \\ \cline{3-7}
          &         &  $5.7 \cdot 10^{-2}$ & $ 4.0 - 8.0 \cdot 10^{-2}$
& 3.3 & 3.2  & 2.9  \\ \cline{3-7}
          &         &  0.11  & $ 8.0 - 16.0 \cdot 10^{-2}$
& 1.5 & 1.4  & 1.2   \\ \hline
 1000     & 640 - 1280 &  $1.4 \cdot 10^{-2}$ & $ 1.0 - 2.0 \cdot 10^{-2}$
& 2.7 & 2.6  & 2.5  \\ \cline{3-7}
          &         &  $2.8 \cdot 10^{-2}$ & $ 2.0 - 4.0 \cdot 10^{-2}$
& 2.5 &2.4  & 2.2  \\ \cline{3-7}
          &         &  $5.7 \cdot 10^{-2}$ & $ 4.0 - 8.0 \cdot 10^{-2}$
& 1.6 & 1.5  & 1.4  \\ \cline{3-7}
          &         &  0.11  & $ 8.0 - 16.0 \cdot 10^{-2}$
& 0.75 & 0.67  & 0.59   \\ \hline
 2000     & 1280 - 2560 &  $2.8 \cdot 10^{-2}$ & $ 2.0 - 4.0 \cdot 10^{-2}$
& 0.90 & 0.87  & 0.83  \\ \cline{3-7}
          &         &  $5.7 \cdot 10^{-2}$ & $ 4.0 - 8.0 \cdot 10^{-2}$
& 0.72 & 0.69  & 0.64  \\ \cline{3-7}
          &         &  0.11  & $ 8.0 - 16.0 \cdot 10^{-2}$
& 0.37 & 0.34  & 0.30   \\ \hline
 5000     & 2560 - 10000 &  0.11  & $ 8.0 - 16.0 \cdot 10^{-2}$
& 0.17 & 0.13  & 0.096   \\ \hline
\end{tabular}
\newpage
%

\newpage
Figure Captions
\begin{description}
\item[Fig.1.]
$h^{(1)}_{A,T,g}(\eta,\xi)$ (eq. (9)) versus $\eta$ for $\xi = 10^{-2}$
(solid line), $\xi = 1$
(dotted line), $\xi = 3.16$ (short-dashed line), $\xi = 31.6$
(long-dashed line) and $\xi = 316$ (dot-dashed line).
\item[Fig.2.]
$h^{(1)}_{F,T,g}(\eta,\xi)$ (eq. (10)) versus $\eta$ for $\xi = 10^{-2}$
(solid line), $\xi = 1$
(dotted line), $\xi = 3.16$ (short-dashed line), $\xi = 31.6$
(long-dashed line) and $\xi = 316$ (dot-dashed line).
\end{description}
\end{document}